\documentclass[preprint2]{aastex}

\newcommand{\kms}{km s$^{-1}\;$}
\newcommand{\kmss}{km s$^{-1}$}
\newcommand{\km}{km s$^{-1}\;$}

\newcommand{\vlsr}{$V$$_{\rm LSR}$}

\newcommand{\sm}{\mbox{$M_{\sun}\;$}}
\newcommand{\lsun}{\mbox{$L_{\sun}$}}

\newcommand{\ho}{H$_{2}$O$\;$}
\newcommand{\hho}{H$_{2}$O}
\newcommand{\mb}{mJy beam$^{-1}$}

\newcommand{\nrest}{$\nu_{\rm {rest}}$}
\shorttitle{321 GHZ \ho MASER IN THE CIRCINUS GALAXY}
\shortauthors{Hagiwara et al.}

\begin{document}


\title{SUBMILLIMETER H$_2$O MASER IN CIRCINUS GALAXY \\
    - A NEW PROBE FOR THE CIRCUMNUCLEAR REGION OF ACTIVE GALACTIC NUCLEI}


\author{Yoshiaki Hagiwara\altaffilmark{1,2}, Makoto Miyoshi\altaffilmark{1,2}, Akihiro Doi\altaffilmark{3}, and Shinji Horiuchi\altaffilmark{4}}

%
\altaffiltext{1}{National Astronomical Observatory of Japan, 2-21-1, Osawa, Mitaka, 
 Tokyo 181-8588, Japan; yoshiaki.hagiwara@nao.ac.jp}
\altaffiltext{2}{Department of Astronomical Science, The Graduate University for 
Advanced Studies (Sokendai), 2-21-1, Osawa, Mitaka, 181-8588 Tokyo, Japan}
\altaffiltext{3}{The Institute of Space and Astronautical Science, Japan Aerospace Exploration Agency, 3-1-1, Yoshinodai, Chuou-ku,  Sagamihara, Kanagawa 252-5210, Japan}
\altaffiltext{4}{CSIRO Astronomy and Space Science, Canberra Deep Space Communications Complex,
PO Box 1035, Tuggeranong, ACT 2901, Australia}
%
\begin{abstract} 
We present the first detection of extragalactic submillimeter \ho maser in the 321 GHz transition toward the center of Circinus galaxy, the nearby Type 2 Seyfert using the Atacama Large Millimeter/Submillimeter Array. We find that Doppler features of the detected 321 GHz \ho maser straddle the systemic velocity of the galaxy as seen in the spectrum of the known 22 GHz 
\ho maser in the galaxy. By comparing the velocities of the maser features in both transitions, it can be deduced that the 321 GHz maser occurs in a region similar to that of the 22 GHz maser, where the sub-parsec-scale distribution of the 22 GHz maser was revealed by earlier very long baseline interferometry observations. The detected maser features remain unresolved at the synthesized beam of $\sim$ $0\arcsec$.66 ($\sim$15 pc) and coincide with the 321 GHz continuum peak within small uncertainties.  We also present a tentative detection of the highest velocity feature (redshifts up to $\sim $635 \kmss) in the galaxy. If this high-velocity feature arises from a Keplerian rotating disk well established in this galaxy, it is located at a radius of $\sim$ 0.018 pc ($\sim$1.2 $\times$ 10$^5$ Schwarzschild radii), which might probe molecular material closest to the central engine.  
\end{abstract}
%
%



\section{INTRODUCTION}
The \ho masers in the 6$_{16}$--5$_{23}$ transition (rest frequency (\nrest) = 22.23508 GHz) in active galactic nuclei (AGNs), what we call megamasers, have been detected in $\sim$ 150 galaxies, most of which are detected exclusively toward narrow-line AGNs having a Seyfert 1.8--2.0 or LINER nucleus \citep[e.g.,][]{kond06,hagi03b,hagi07}. Some of these megamasers show spectra giving evidence for emission from edge-on rotating disks on scales of sub-parsecs as measured in \object{NGC\,4258} with very long baseline interferometry (VLBI) observations at milliarcsecond angular resolution \citep[e.g.,][]{miyo95,hagi01,kuo11}. According to theoretical studies, the 22 GHz \ho maser is inverted in wider range of physical conditions at kinetic temperature of $T_k$ = 200--2000 K, hydrogen densities of $n$(H$_2$) = 10$^{8}$--10$^{10}$ cm$^{-3}$, and \ho 
densities of $n$(H$_2$O) = 10$^3$--10$^5$ cm$^{-3}$, {which largely excites submillimeter \ho maser transitions such as the 321 GHz (10$_{29}$--9$_{36}$) and 325 GHz (5$_{15}$--4$_{22}$) transitions} \citep{deg77,neu90,neu91,yat97}. These studies have shown that the 321 GHz maser is even strongly excited under more restricted physical conditions of $T_k$ $>$ 1000 K. The upper state energy of the 321 GHz transition is higher than the 22 GHz transition ($E_u/k$=1862 K for the 321 GHz transition and $E_u/k$=644 K for the 22 GHz transition).

The first discovery of extragalactic sub-millimeter \ho masers was achieved by \citet{liz05}, in which the detection of \ho masers in the 183 GHz (3$_{13}$--2$_{20}$) and 439 GHz (6$_{43}$--5$_{50}$) transitions was reported in the nuclear region of \object{NGC\,3079}, the Type 2 Seyfert/LINER galaxy, although the detection of the 439 GHz maser was tentative. The \ho emission in the 183 GHz transition was also detected toward the luminous infrared galaxy \object{Arp220} \citep{cer06}. However, the detected emission could be thermal emission.

\object{Circinus galaxy} at a distance of 4.2 $\pm$ 0.8 Mpc is a nearby spiral lying behind the Galactic plane\citep{free77}, exhibiting evidence of the co-existence of starbursts occurring within a nuclear molecular ring in CO emission \citep{cur08}. The galaxy hosts a Type 2 Seyfert nucleus having a luminous ($\sim$ 10$^{42}$ erg~s$^{-1}$) hard X-ray source obscured by Compton-thick medium \cite{mat96}. 
The adopted systemic velocity of the galaxy is 433 \kms \citep{dev91,bra03}. (The optical velocity definition is adopted throughout this article and the velocities are calculated with respect to the local standard of rest (LSR).)

{Bright 22 GHz \ho maser emission in the galaxy was first
reported by \citet{gard82} and the known highly variable flux density of the maser is in the range of $\sim$5 to $\sim$40Jy} \citep[e.g.,][]{linc97,bra03,jam05}.
 {The velocity range of highly Doppler-shifted (high-velocity) maser features detected to date is $\sim$50 to $\sim$900 \kmss} \citep{linc03b}.
Follow-up VLBI observations of the 22 GHz maser at the milliarcsecond resolution revealed that the maser traces a warped edge-on disk that implies a sub-Keplerian rotation with a radius of $\sim$0.1--0.4 pc at the maximum rotation velocity of $\sim$270 \kms around a nucleus and outflowing components distributed near the disk with the velocity of $\sim$160 \kms \citep{linc03a}. 
 The Circinus galaxy, already known to show the strongest 
flux density of the 22 GHz \ho maser among the \ho megamasers, could show the 321 GHz \ho emission as \ho masers in both the 22 and 321 GHz transitions
have been detected in the same Galactic sources \citep{men90,men91}. \\
In this work, we present the observation of the 321 GHz \ho maser emission 
in Circinus galaxy, obtained using Atacama Large Millimeter/submillimeter Array (ALMA) in the Cycle 0 program. 
\section{OBSERVATIONS AND DATA REDUCTION}
We observed \ho emission in the 10$_{29}$--9$_{36}$ transition (\nrest=321.226 GHz) toward the \object{Circinus galaxy} on 2012 June 3 using the ALMA as part of a search for extragalactic submillimeter \ho maser. 
Observations lasted for 49 minutes (approximately 25 minutes on-source) with 18 antennas. The tracking center of the galaxy was $\alpha$(J2000) = 14$^{\rm h}$13$^{\rm m}$09$^{\rm s}$.906, $\delta$(J2000) =
 $-$65$\degr$20$\arcmin$20$\arcsec$.468.
These observations were made with a single dual polarization spectral window (1.875 GHz bandwidth), divided into 3840 spectral channels, yielding spectral resolutions of 488.3 kHz or 0.458 \kms at the observed
frequency of 319.4 GHz. A resultant total velocity coverage is $\sim$ 1760 \kmss. The 1.875 GHz spectral window was centered near the galaxy's systemic velocity of 433 \kmss.
Phase-referencing observations were conducted, by switching to a phase-referencing source, J\,1329--5608, 
the position of which is $\approx$ 9 $\degr$ away from the target source.
Data calibration was performed using the Common Astronomy
Software Applications (CASA).
Amplitude calibration was performed using observations of Titan, 
and the bandpass correction was made with 3C\,279. 
Flux-density errors of 10 \% are conservatively adopted, based on 
the capabilities of Cycle 0 observations in Band 7.
Imaging was performed using CASA in natural weighting, for which the synthesized beam size was 0$\arcsec.66 \times 0\arcsec$.51  (P.A.= --17$\degr$). The image analysis was performed using both the CASA and the Astronomical Image Processing Software.
After the phase and amplitude calibrations, the 321 GHz continuum emission of the Circinus was subtracted from the spectral line visibilities using line-free channels prior to the imaging and CLEAN deconvolution of line
emission. The line emission in the Circinus galaxy was thus separated out from the continuum emission. The rms noise level of spectral line maps in natural weight was $\sim$9--11 \mb~ per spectral channel, depending on each frequency channel. The rms of the continuum map was $\sim$0.8 \mb.

\section{RESULTS}
\begin{figure}
\epsscale{1.0}
\plotone{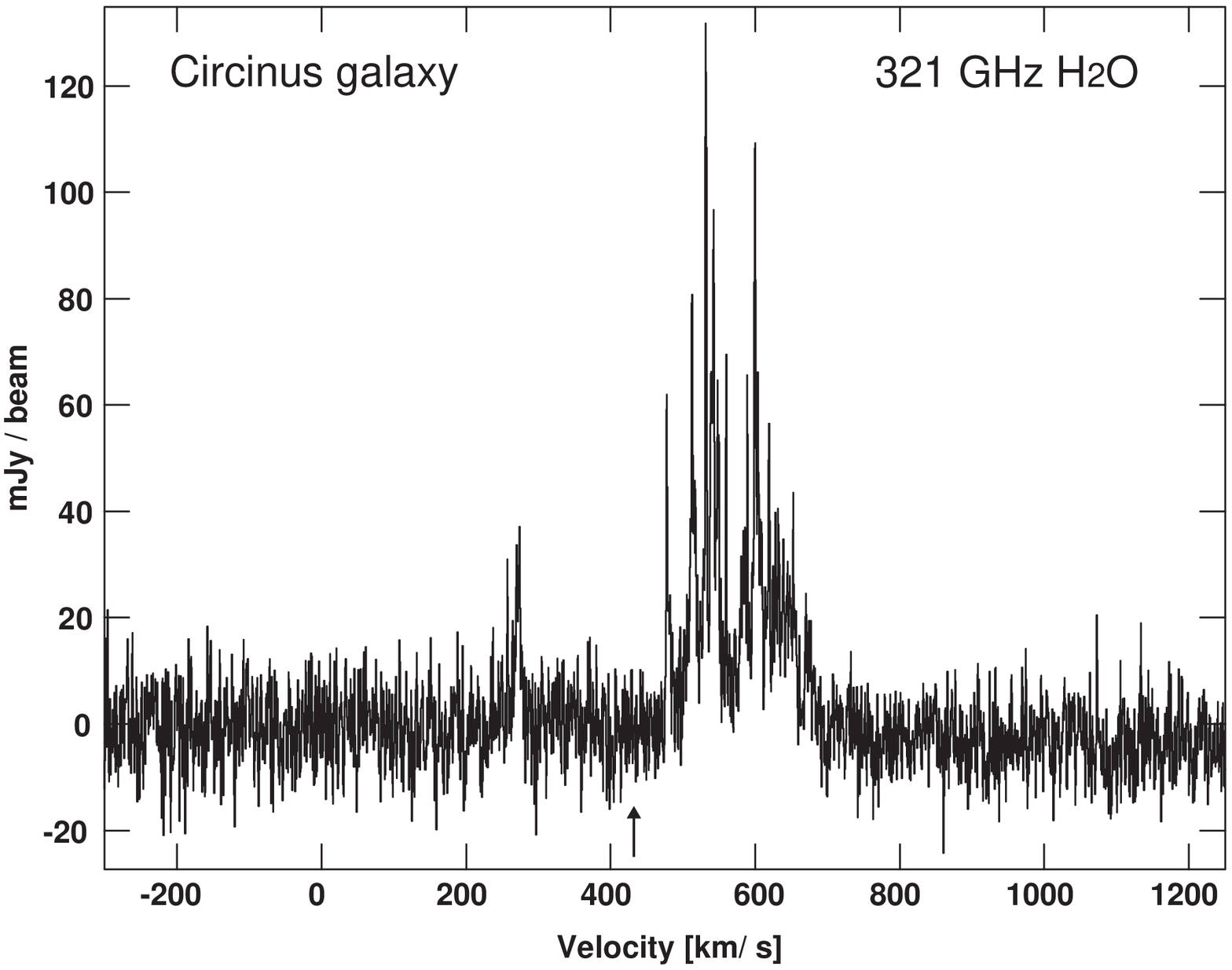}
\caption{Spectrum of the 321 GHz \ho maser from the center of the Circinus galaxy at the spectral resolution of 488.3 kHz or 0.458 \kmss, obtained using 18 antennas on 2012 June 3 in the Cycle 0 ALMA program.  The total velocity range covered in the spectrum is \vlsr=--300.0--1250.0 \kmss, outside of which no significant emission was detected. The LSR systemic velocity, 433 \kms is denoted by an arrow. Amplitude scale is in \mb. 
\label{fig1}}
\end{figure}

\begin{figure}
\epsscale{1.0}
\plotone{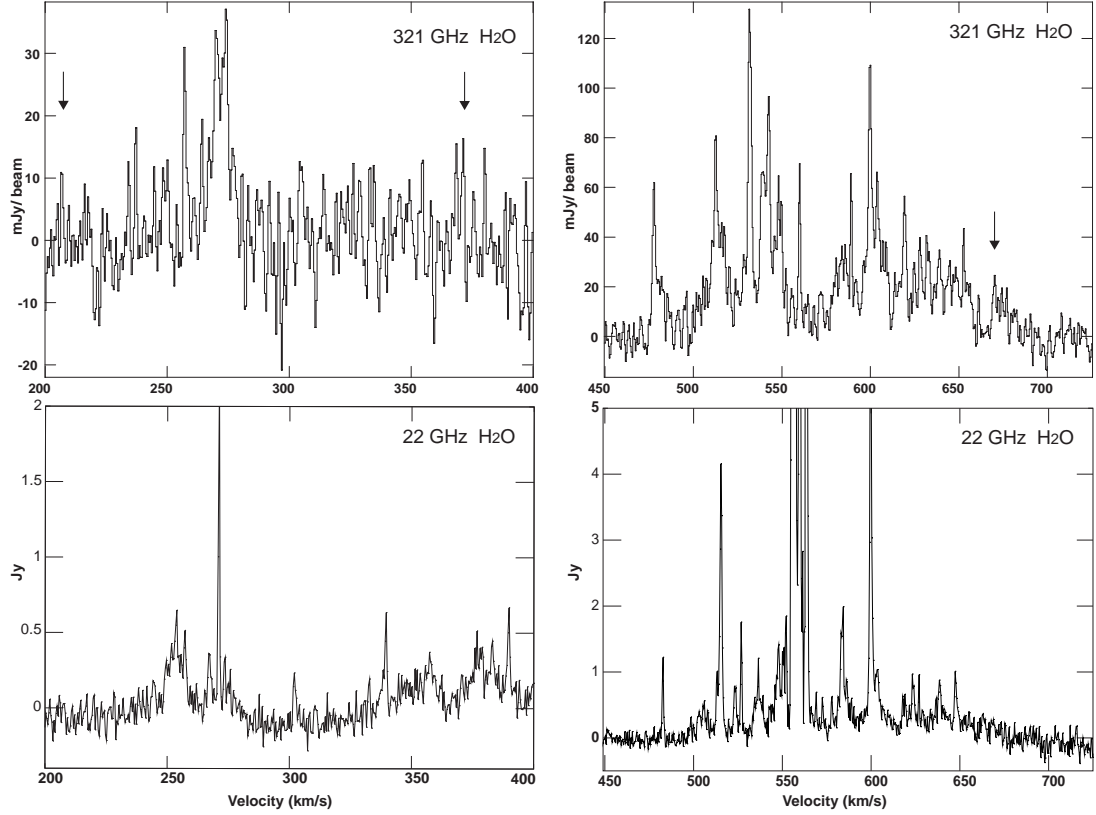}
\caption{Comparison of the \ho spectra between lines in the 321 GHz and 22 GHz transition toward the center of the Circinus galaxy. The 22 GHz spectra were obtained with the Deep Space Network (DSN) 70 m antenna located at Tidbinbilla (S. Horiuchi et al., in preparation). The observations of the Tidbinbilla 70 m were made on 2012 September 7 at the spectral resolution of 31.25 kHz or 0.42 \kms with an rms of $\sim$2.5 mJy and the 321 GHz spectra were on 2012 June 3. Minor features (\vlsr=206.4, 370.9, and 668.7~\kmss) are denoted by arrows. The flux density scales of the 22 GHz spectra shown in the figures are rather tentative and require detailed analysis (S. Horiuchi et al., in preparation). Upper left: blueshifted velocity range (200--400 \kmss) of the 321 GHz maser spectrum obtained from Figure\ref{fig1}. Lower left: spectrum of the blueshifted maser obtained at Tidbinbilla. Upper right: redshifted velocity range (450--725 \kmss) of the spectrum from Figure\ref{fig1}. Lower right: spectrum of the redshifted maser obtained at Tidbinbilla. \label{fig2}}
\end{figure}

\begin{figure}
\epsscale{1.0}
\plotone{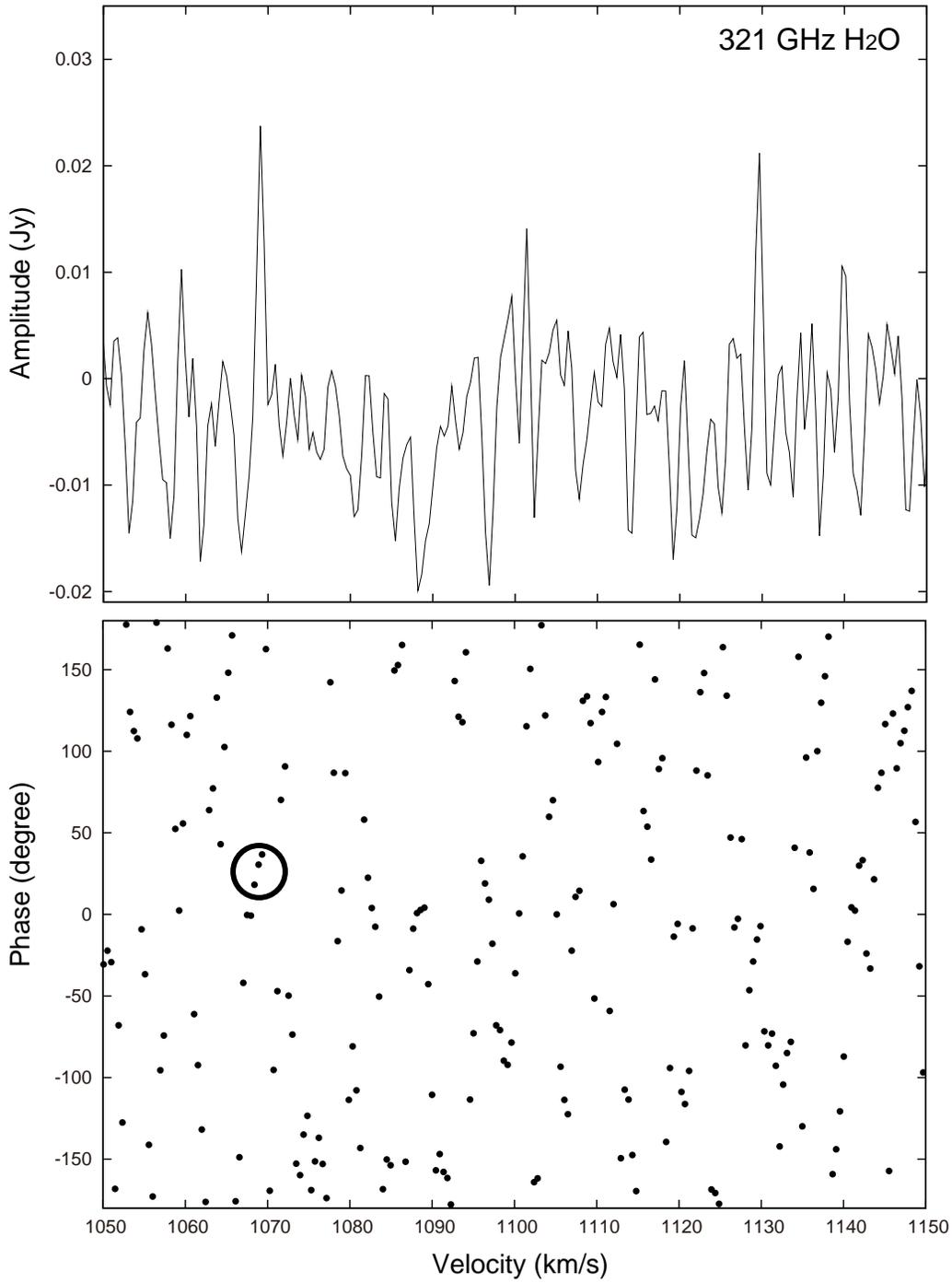}
\caption{Redshifted 321 GHz maser spectrum: a highly redshifted feature tentatively detected ($\sim$ 3$\sigma$ ) peaking at \vlsr= 1069 \kms and a possible detection of a feature at \vlsr= 1129.7 \kmss. Upper: amplitude of the spectrum between \vlsr= 1050 and 1150 \kmss. Lower: corresponding phase. The phases of the 1069 \kms feature are indicated by a circle.
\label{fig3}}
\end{figure}

The 321 GHz \ho maser in the Circinus galaxy was searched within the nominal 
field of view of 18\arcsec, centered on the phase-tracking center of the galaxy and in the velocity range of \vlsr=$-$430 to $+$1290 \kmss, covering the total velocity range of known 22 GHz \ho maser measured in earlier single-dish observations \citep{gard82,naka95,linc97,bra03,jam05}. We detected the 321 GHz \ho line emission, the peak flux density of which is 131 \mb~ at \vlsr=531.1 \kms with a signal-to-noise ratio (S/N) of $>$10 (Figure~\ref{fig1}) and the total flux density is $\sim$350 mJy. The total integrated intensity estimated from all the detected \ho emission is $\sim$14 Jy \kmss. The apparent luminosity of the emission is $\sim$ 5~$\lsun$, assuming isotropic radiation of the emission.
The detected \ho emission can be divided into two groups, one of which is blueshifted emission of \vlsr=$\sim$255--$\sim$275 \kms peaking at \vlsr=274.4 \kms and the other is redshifted emission lying from \vlsr=$\sim$470--$\sim$670 \kmss. The line shapes of these Doppler-shifted features are narrow, similar to the \ho maser features seen in the 22 GHz single-dish spectra (Figure~\ref{fig2}). 
The redshifted emission is composed of broad features, narrow-line features peaking at \vlsr=477.1, 511.4, 530.6, 531.1, 541.6, 547.1, 559.0, 587.8, 598.8, 602.4, 618.0 \kmss, and a minor peak at \vlsr= 668.7 \kms  (denoted by an arrow in Figure~\ref{fig2}). Figure~\ref{fig3} displays a highly redshifted feature  tentatively detected in $\sim$ 3$\sigma$ level centered at \vlsr=1069.1 \kmss.  
One may suggest a possible detection of a feature peaking at \vlsr= 1129.7 \kmss ($<$ 3$\sigma$), which must be confirmed in further measurements.

\begin{figure}
\epsscale{1.0}
\plotone{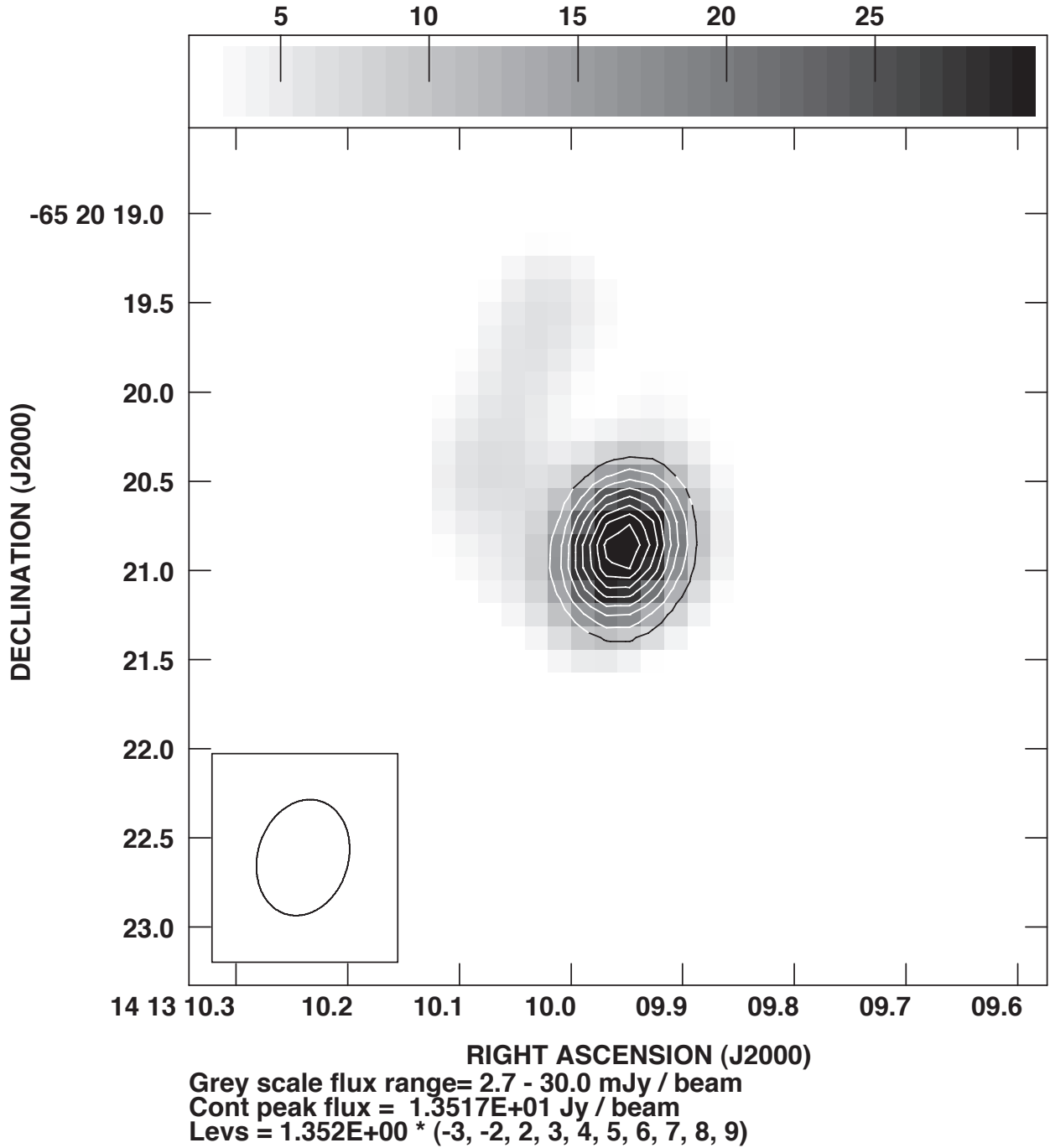}
\caption{Integrated intensity image (contours) of the 321 GHz \ho maser overlaid the 321 GHz continuum image (gray scale) observed with the ALMA. The peak integrated intensity of the \ho maser and the peak flux density of the continuum are 13.6 Jy \km beam$^{-1}$ and 40.6 \mb, respectively. Contour levels and gray scales (scales are from 2.7 mJy ($\sim$3$\sigma$) to 30 mJy) are denoted in the image.
The synthesized beam (FWHM) is plotted in the left bottom corner.
\label{fig4}}
\end{figure}

In Figure~\ref{fig2}, we made a comparison of the 321 GHz \ho spectra with 22 GHz spectra obtained on 2012 September 7 using the Deep Space Network 70 m antenna at Tidbinbilla. 
The known strong variability of the 22 GHz maser in this galaxy makes it difficult to compare precisely the velocities of each feature between the 22 GHz and 321 GHz lines. In our 321 GHz observation, no feature is detected at or near the systemic velocity, consistent with the fact that earlier observations detected no distinct systemic feature of the 22 GHz maser in the galaxy \citep[e.g.,][]{linc03b,bra03}. It should be also remarked that the asymmetry between the blueshifted and redshifted emission is observed as in the cases of other \ho megamasers such as NGC 4258: Redshifted features are more numerous and are significantly more intense than the blueshifted features \citep[e.g.,][]{mao98}. 
Thus, the 321 GHz emission in the galaxy largely overlap Doppler features appearing in the 22 GHz maser spectra obtained in the earlier single-dish observations. The Gaussian-fitted position (relative to the phase-referencing source) of the redshifted peak emission at 531.6 \km is $\alpha$(2000):14$^{\rm h}$13$^{\rm m}$09$\fs$95$\pm$0.01, $\delta$(2000): --65$\degr20\arcmin 20\arcsec.92$$\pm$0.02. 
The position of the blueshifted peak at 274.2 \kms is  \\
$\alpha$(2000):14$^{\rm h}$13$^{\rm m}$09$\fs$96$\pm$0.03, $\delta$(2000): -65$\degr20\arcmin 20\arcsec$.89 $\pm$ 0.03.  
 The relative positions of other redshifted and blueshifted features coincide with these positions within uncertainties of fitting errors and these features remain unresolved at the angular resolution of $\sim$ 0$\arcsec$.66, or 14.6 pc. \\
The 321 GHz submillimeter continuum was also detected in the center of the galaxy with an S/N of $>$ 50 (Figure~\ref{fig4}), the peak position of which is $\alpha$(2000):14$^{\rm h}$13$^{\rm m}$09.96$\pm$0.01, $\delta$(2000): --65$\degr20\arcmin 20.87\arcsec$$\pm$0.01. The detected nuclear continuum with the total flux density of 55 mJy was partly resolved in the northeast direction accompanying an elongated substructure in the north (Figure~\ref{fig4}).
According to the ALMA Cycle 0 Capabilities, the positional errors 
are $\sim$ 10\% of the synthesize beam, 0$\arcsec$.66/10 or 0$\arcsec$.066 in our observation, which seems to be the most dominant source of an error. Approximate relative positional errors between the \ho emission and the continuum peak, represented by the synthesized beam size divided by 2 $\times$ S/N \citep[e.g.,][]{hagi03b} are $\sim$ 0$\arcsec$.07 or 1.5 pc. 
Consequently, the relative positions between the maser and continuum emission peak are consistent within these errors. Note that the positions of the blueshifted and redshifted maser peaks at \vlsr=274.2 and 531.6 \km are offset from the optical nucleus at $\alpha$(2000):14$^{\rm h}$13$^{\rm m}$09.950$\pm$0.335, $\delta$(2000): --65$\degr20\arcmin 21.20\arcsec$$\pm$0.25 (primary from NED) by --0$\arcsec$.28 to --0$\arcsec$.31 in declination, although these offsets are almost within errors.
\\
The resolved 321 GHz continuum emission shows an elongated structure (Figure~\ref{fig4}),
the direction of which is similar to that of a molecular or atomic gas disk at an inclination of $\sim$ 60$\degr$--70$\degr$ \citep{cur08}.
\section{DISCUSSION}
\subsection{Origin of the 321 GHz \ho and Submillimeter Continuum Emission}
The synthesized beam size of our observation was too large to determine if the detected 321 GHz \ho emission is thermal emission or maser, from the value of brightness temperature.
Given that the 321 GHz line features are confined within a diameter of $\la$14 pc and the line shapes are narrow  ($<$0.45 \kmss), it is most likely that the \ho emission detected in the Circinus galaxy is due to maser amplification. \\
To investigate the nature of the 321 GHz continuum, more information such as a spectral index would be required. We can speculate about the possibility of thermal emission from dust as the continuum shows an extended part in the general direction of a galactic-scale disk in a dusty environment, while a central nuclear continuum could contain the contribution of free--free emission.

Since the peak position of the continuum coincides with that of the maser, the maser could be enhanced through the thick dust lane in our line of sight.
The optical studies revealed that the central 15 pc of the galaxy is dominated by the AGN luminosity, with star formation
contributing only 2\% \citep{sos01}. The 321 GHz maser in the Circinus galaxy could be associated with AGN activity. Because of the relatively smaller maser luminosity, we still cannot still rule out that the maser originates in star-forming activity.
The distinct velocity gradient of the 321 GHz maser
is not detected throughout our analysis of the first moment map produced from the spectral-line cube containing features lying at \vlsr=256--670 \kmss, suggesting that the maser emission is neither spatially nor kinematically resolved in our observation on scales of 10 pc, while, according to the earlier VLBI observations, the 22 GHz maser in the galaxy was resolved on scales of $\approx$1 pc \citep{linc03b,jam09}, implying that a better angular resolution is required, at least, better by one order of magnitude than that of our observation.  The future VLBI mapping of the 321 GHz maser at milliarcsecond resolution should be able to clarify sub-parsec-scale kinematics of the galaxy by measuring distributions of each maser spot.
\subsection{Comparison of 321 GHz \ho Emission with 22 GHz \ho Maser}
In Figure~2, one can compare the spectra of the 321 GHz maser emission 
with the 22 GHz maser emission, both of which were observed at approximately the same time in 2012 June and September. The velocity ranges of masers at both transitions overlap to some extent and some of the prominent maser features peaked at or near the same velocities (see Figure 2). However, the velocity span of the 321 GHz maser is smaller than that of the 22 GHz maser due to the spectrum having a broader velocity coverage \citep{linc03b}; high-velocity features of the 22 GHz transition span from $\sim$50 \kms to $\sim$900 \km having the most Doppler-shifted velocity of 460 \kms with respect to the systemic velocity \citep{linc03b}, while those of the 321 GHz transition from $\sim$270 \kms to $\sim$670 \kms in our observation, excepting the minor detection. This 
might be explained by the fact that the 321 GHz maser is present in 
more physically restricted regions in higher temperature and density as predicted in theoretical studies \citep{neu90,yat97}, which should be examined by resolving the spatial distribution of the maser at higher angular resolution. However, with the comparable sensitivity 
of the earlier 22 GHz maser spectra, the total velocity range of the 321 GHz maser emission might be as broad as that of the 22 GHz maser. In the case of the Cepheus A star-forming regions, the 321 GHz and 22 GHz masers were observed within the region of $\sim$ 1$\arcsec$, in which spatial coincidence between the 321 GHz and 22 GHz masers was not found and the 321 GHz maser is believed to be tracing warmer gas than the 22 GHz maser \citep{pat07}.  We note that the 321 GHz maser in evolved stars is highly variable, compared with 22 GHz and 325 GHz masers \citep{yat96}. Thus, the study of the intensity variability of the 321 GHz maser in the galaxy is of great interest. \\

From Figure 2, it is difficult to find one-to-one correspondence of each velocity feature between the two transitions. However, the overall velocity spread of the 321 GHz maser features looks similar to that of 22 GHz masers that are believed to trace a masing disk rotating around a central engine having 1.7 $\times$ 10$^6$ \sm enclosed mass and 
non-disk maser components from bipolar outflows lying within about 1 pc from the edge-on disk \citep{linc03a}. At the angular resolution on our map, it is not possible to distinguish only from their velocities whether or not each of the detected 321 GHz maser features is coming from the disk or non-disk outflows. 
%
%
\subsection{Possible Detection of High-velocity Gas in the Circinus}
 The known highest velocity maser feature 
in Circinus galaxy is redshifted by $\sim$460 \kms from the systemic velocity \citep{linc03b}, while the blueshifted counterpart was not detected in our 321 GHz spectra. Assuming that the 1069 \kms feature (redshifts up to $\sim$ 635 \kmss) is real, it is located at a radius of $\sim$ 0.018 pc ($\sim$ 1.2 $\times$ 10$^5$ Schwarzschild radii for a 1.7 $\times$ 10$^6$ \sm black hole)
from the central engine, by calculation based on a Keplerian ($v$ $\propto$ $r^{-0.5}$) rotation disk model in Greenhill et al. (2003a, 2003b). This high-velocity feature could probe the inner most part of a molecular gas disk in the galaxy. The submillimeter maser is potentially a powerful tool for exploring regions which cannot be probed with the 22 GHz maser. \\

The interpretation of the origin of the \ho maser in the Circinus galaxy is not straightforward. Nevertheless, our discovery of the 321 GHz \ho maser in Circinus galaxy will be useful as a basis for future detailed studies of circumnuclear gas in AGNs. The detected 321 GHz and well-studied 22 GHz traditions are from ortho-\hho.
Using these two different maser transitions will enable us to place new constraints on radiative transfer models \citep{liz07} to determine physical conditions of molecular gas in the inner sub-parsecs of AGNs.
Moreover, the \ho lines at 22 and 321 GHz show higher atmospheric transmission 
than those at 183 and 325 GHz, which is advantage for interferometric observations.
Ultimately, we hope to measure the angular distribution of submillimeter \ho masers including those of 
the 321 GHz or other transitions on the basis of currently on-going efforts in developments of VLBI at submillimeter wavelengths.

\acknowledgments
This Letter makes use of the following ALMA data: 2011.0.00121.S. ALMA is a partnership of ESO (representing its member states), NSF (USA) and NINS (Japan), together with NRC (Canada) and NSC and ASIAA (Taiwan), in cooperation with the Republic of Chile. The Joint ALMA Observatory is operated by ESO, AUI/NRAO and NAOJ. The Tidbinbilla 70m telescope is part of the NASA Deep Space Network and is operated by CSIRO. This research has made use of the NASA/IPAC Extragalactic Database (NED) which is operated by the Jet Propulsion Laboratory, California Institute of Technology, under contract with the National Aeronautics and Space Administration. Finally, we thank the anonymous referee for helpful suggestions that significantly improved the article. 
%


\onecolumn



\begin{thebibliography}{}
%
\bibitem[Braatz et al.(2003)]{bra03}Braatz, J. A., Wilson, A. S., Henkel, C., Gough, R., \& Sinclair, M. 2003, \apjs, 146, 249
%
\bibitem[Cernicharo et al.(2006)]{cer06}Cernicharo, J., Pardo, J. R., \& Weiss, A. 2006, ApJL, 646, L49
%
\bibitem[Curran et al.(2008)]{cur08}Curran, S. J., Koribalski, B. S., \& Bains, I. 2008, \mnras, 389, 63
%
\bibitem[Deguchi(1977)]{deg77} Deguchi, S. 1977, \pasj, 29, 669
%
\bibitem[de Vaucouleurs et al.(1991)]{dev91} de Vaucouleurs, G., de Vaucouleurs, A., Corwin, H. G., et al. 1991, Third Reference Catalog of Bright Galaxies (Berlin: Springer)
%
\bibitem[Freeman et al.(1977)]{free77}Freeman, K. C., Karlsson, B., Lynga, G., et al. 1977, \aap, 55, 445
%
%
\bibitem[Gardner \& Whiteoak(1982)]{gard82} Gardner, F. F., \& Whiteoak, J. B. 1982, \mnras, 201, P13
%
%
\bibitem[Greenhill et al.(2003a)]{linc03a} Greenhill, L. J., Booth, R. S., Ellingsen, S. P., et al. 
 2003a, \apj, 590, 162
%
\bibitem[Greenhill et al.(1997)]{linc97} Greenhill, L. J., Ellingsen, S. P., Norris, R. P., et al. 1997, ApJL, 474, L103
%
%
\bibitem[Greenhill et al.(2003b)]{linc03b} Greenhill, L. J., Kondratko, P. T., Lovell, J. E. J., et al. 2003b, ApJL, 582, L11
%
\bibitem[Hagiwara(2007)]{hagi07} Hagiwara, Y. 2007, \aj, 133, 1176
%
\bibitem[Hagiwara et al.(2003a)]{hagi03a} Hagiwara, Y., Diamond, P. J., \& Miyoshi, M. 2003a, \aap, 400, 457
%
\bibitem[Hagiwara et al.(2003b)]{hagi03b}
 Hagiwara, Y., Diamond, P. J., Miyoshi, M., Rovilos, E., \& Baan, W. 2003b, \mnras, 344, L53
%
\bibitem[Hagiwara et al.(2001)]{hagi01} Hagiwara, Y., Diamond, P. J., Nakai, N., \& Kawabe, R. 2001, \apj, 560, 119
%
\bibitem[Humphreys(2007)]{liz07}  Humphreys, E. M. L. 2007, in IAU Symp. 242, Astrophysical Masers and their Environments, ed. J. M. Chapman \& W. A. Baan (Cambridge: Cambridge Univ. Press), 471 
%
\bibitem[Humphreys et al.(2005)]{liz05} Humphreys, E. M. L., Greenhill, L. J., Reid, M. J., et al. 2005, ApJL, 634, L133
%
\bibitem[Kondratko et al.(2006)]{kond06} Kondratko , P. T., Greenhill, L. J., Moran, J. M., et al. 2006, \apj, 638, 100
%
\bibitem[Kuo et al.(2011)]{kuo11} Kuo, C. Y., Braatz, J. A., Condon, J. J., et al. 2011, \apj, 727, 20
%
\bibitem[Maoz \& McKee (1998)]{mao98}Maoz, E., \& McKee, C. F. 1998, \apj, 494, 218
%
\bibitem[Matt et al.(1996)]{mat96} Matt, G., Fiore, F., Perola, G. C., et al. 1996, \mnras, 281, L69
\bibitem[McCallum et al.(2005)]{jam05} McCallum, J. N.,  Ellingsen, S. P., Jauncey, D. L., Lovell, J. E. J., \& Greenhill, L. J. 2005, \aj, 129, 1231
%
\bibitem[McCallum et al.(2009)]{jam09} McCallum, J. N., Ellingsen, S. P., Lovell, J. E. J., Phillips, C. J., \& Reynolds, J. E. 2009, \mnras, 392, 1339
%
%
\bibitem[Menten \& Melnick(1991)]{men91} Menten, K. M. \& Melnick G. J. 1991, \apj, 377, 647
%
\bibitem[Menten et al.(1990)]{men90} Menten, K. M., Melnick G. J., \& Phillips, T. G. 1990, ApJL, 350, L41
%
\bibitem[Miyoshi et al.(1995)]{miyo95}
 Miyoshi, M.  Moran, J., Herrnstein, J., et al. 1995, Natur, 373, 127
%
\bibitem[Nakai et al.(1995)]{naka95} Nakai, N., Inoue, M., Miyazawa, K., Miyoshi, M., \& Hall, P. 1995, PASJ, 47, 771
%
\bibitem[{{Neufeld} \& {Melnick}(1990)}]{neu90}
{Neufeld}, D.~A. \& {Melnick}, G.~J. 1990, ApJL, 352, L9
%
\bibitem[Neufeld \& Melnick(1991)]{neu91}
{Neufeld}, D.~A. \& {Melnick}, G.~J. 1991, \apj, 368, 215
%
\bibitem[Patel et al.(2007)]{pat07}
Patel, N. A., Curial, S., Zhang, Q., et al. 2007, ApJL, 658, L55
%
\bibitem[Sosa-Brito et al.(2001)]{sos01}
Sosa-Brito, R. M., Tacconi-Garman, L. E., Lehnert, M. D. \& Gallimore, J. F. 2001, \apjs, 136, 61
%
\bibitem[Yates et al.(1996)]{yat96}
Yates, J. A. \& Cohen, R. J. 1996, \mnras, 278, 655
%
\bibitem[Yates et al.(1997)]{yat97}
Yates, J. A., Field, D., \& Gray, M. D. 1997, \mnras, 285, 303
\end{thebibliography}
\end{document}